\documentclass[reprint,aip,jcp,showpacs,superscriptaddress,floatfix,letterpaper]{revtex4-1}
\usepackage{amsmath,amssymb,amsbsy,amsfonts,amsthm}
\usepackage{bm}
\usepackage{color}
\usepackage{units}
\usepackage{graphicx}
\setkeys{Gin}{width=\linewidth,totalheight=\textheight,keepaspectratio}
\usepackage{tikz}
\usetikzlibrary{calc,arrows}

\newcommand{\der}[2]{\frac{\displaystyle d #1}{\displaystyle d #2}}

\def\date{}

\begin{document}

\title{Measuring disorder in irreversible decay processes}

\author{Shane~W.~Flynn}
\author{Helen~C.~Zhao}
\author{Jason~R.~Green}
\email[]{jason.green@umb.edu}
\affiliation{Department of Chemistry,\
  University of Massachusetts Boston,\
  Boston, MA 02125
}




\begin{abstract}

Rate coefficients can fluctuate in statically and dynamically disordered
kinetics. Here we relate the rate coefficient for an irreversibly decaying
population to the Fisher information. From this relationship we define kinetic
versions of statistical-length squared and divergence that measure cumulative
fluctuations in the rate coefficient. We show the difference between these
kinetic quantities measures the amount of disorder, and is zero when the rate
coefficient is temporally and spatially unique.

\end{abstract}


\maketitle

\section{Introduction}

Rate coefficients are necessary and sufficient information in the rate laws of
kinetic phenomena. To predict the average behavior of populations with these
laws, it is essential to reliably estimate the rate coefficient. When
fluctuations away from the average are significant in irreversible decay,
predictions depend on a distribution of rate coefficients or a time varying
rate coefficient~\cite{Ross08,Plonka01}. The terms static and dynamic disorder
describe these theoretical constructs and the underlying kinetic
process~\cite{Zwanzig90}. Examples of disordered kinetics include enzyme
catalyzed reactions~\cite{TerentyevaERKHLB12,EnglishMOLLSCKX05,MinELCKX05}, the
dissociation of single molecules during pulling
experiments~\cite{Kuo22062010,ChatterjeeC11}, and hydrogen bond
breaking~\cite{LuzarC96b}. However, static and dynamic disorder are not
mutually exclusive classifications of a process~\cite{Ross08} or
straightforward to assign \textit{a priori}~\cite{Dewey92}, which raises
fundamental and practical questions. How can we measure disorder from kinetic
data when microscopic heterogeneity manifests in the rate coefficient? Or, when
is the rate coefficient well defined? When is traditional kinetics valid?

Answering these questions requires the ability to measure disorder. An example
of recent progress is an information theoretic measure of dynamic
disorder~\cite{LiK13}. Such measures could enable new methods for minimizing
the statistical description of rate coefficients and maximizing their
predictive fidelity for data collected from simulations and experiments. In
this work we demonstrate a theoretical framework for this purpose that applies
to irreversible decay. Central to the theory is a measure of temporal and
spatial fluctuations in the rate coefficient, when the disorder is static,
dynamic, or both. Our main result is an inequality that only reduces to an
equality in the absence of disorder. This result is a necessary condition for
the traditional kinetics of irreversible first-order decay to be valid and the
associated rate coefficients to be unique.

Consider a population of species irreversibly decaying over time. The survival
probability, $S(t)$, is the probability the initial population survives up to a
time $t$. This probability defines the observed or effective rate coefficient,
$k(t)$, through the differential rate law
\begin{equation}
  \label{eqn:key}
  k(t) \equiv -\frac{d\ln S(t)}{dt},
\end{equation}
characterizing the first-order decay of the entire population. If individual
members of the population are experimentally indistinguishable and the overall
decay of $S(t)$ is non-exponential~\cite{WangW94,WangW93,WangW95}, a disordered
model for $k(t)$ may be necessary. Non-exponential decay can imply an underlying
mechanism where members decay in different structural or energetic
environments, or a local environment that fluctuates in time~\cite{WangW95}.
For irreversible decay, the survival probability is the ratio of the number of
members in the population at a time $t$, $N(t)$, and the number initially,
$N(0)$. In terms of the effective rate coefficient it is
\begin{equation}
 S(t) = \frac{N(t)}{N(0)} = e^{-\int_{0}^{t}{k(t')dt'}}.
\end{equation}
Survival probabilities are the input to the theory here and a common
observable; they are measurable from computer simulations and experiments.
While measurements of $S(t)$ may have non-negligible statistical fluctuations,
we use non-fluctuating survival probabilities to simplify the presentation of
our theoretical results.

\section{Theory}

One purpose of the theory here is to measure the fluctuations in the
statistical parameter $k(t)$ for statically and dynamically disordered
kinetics. For activated escape on an energy surface, measures of the
fluctuations in $k(t)$ characterize the mechanism of decay over a distribution
of barriers or a single barrier whose height varies in time. Our approach is to
use the Fisher information, which is a natural measure of the ability to
estimate statistical parameters from the probability distribution of a
fluctuating observable. Colloquially, the Fisher information is the amount of
``disorder'' in a system or phenomenon~\cite{Frieden04}, but is it the amount
of static and dynamic disorder in a kinetic process? To answer this question, a
relationship between the Fisher information and rate coefficients, the
parameters in kinetic phenomena, is necessary.

We relate $k(t)$ and a modified form of the Fisher information, expressed in
terms of the survival probability of a decaying population through
Equation~\ref{eqn:key}. Assuming an exact, non-fluctuating survival probability
varying smoothly over time describes the evolution of the population, a
reasonable form of the Fisher information at time $t$ is
\begin{equation}
  \label{eqn:fisherinfo}
  I(t) \equiv S(t)\left(\der{\ln S(t)}{t}\right)^2 = k(t)^2S(t),
\end{equation}
which shows the direct relation to the effective rate coefficient. A more
general form of the Fisher information may be necessary when there are
statistical fluctuations in $S(t)$~\cite{Frieden04}. Rearranging
Equation~\ref{eqn:fisherinfo} gives
\begin{equation}
  \label{eqn:fisherinrate}
  k(t) = \sqrt{I(t)/S(t)}.
\end{equation}
The remainder of our results stem from this connection between the Fisher
information and the statistical parameter, $k(t)$, of interest in disordered
kinetics.

A particularly useful aspect of the Fisher information, in general, is that it
naturally defines a ``statistical distance'' between probability distributions.
Though the notion of statistical distance was originally applied to quantum
states~\cite{Wootters81,BraunsteinC94}, it is applicable to any two probability
distributions. With the definition of the Fisher information above, we can
measure the statistical distance between survival probabilities and relate it
to the observed rate coefficient. If we take a microscopic, stochastic
perspective, the rate coefficient of decay, $k(t)$, is the transition
probability per unit time. In a given time interval between $t$ and $t+\delta
t$, it is the ratio of the number of reacted molecules $N(0)\delta S\equiv
N(0)[S(t)-S(t+\delta t)]$ and the number of unreacted molecules $N(0)S(t)$
\begin{equation}
  k(t) \equiv \frac{\delta S}{S(t)\delta t}.
\end{equation}
From this discrete form of Equation~\ref{eqn:key} and the discrete form of the
relation between the time-dependent rate coefficient and the Fisher information
in Equation~\ref{eqn:fisherinrate}, we define a dimensionless statistical
distance
\begin{equation}
  \delta s^2 \equiv k(t)^2\delta t^2 = \frac{I(t)}{S(t)}\delta t^2
  = \left[\frac{\delta S}{S(t)}\right]^2 = \left[\delta \ln S\right]^2
\end{equation}
between the logarithm of the survival probabilities at $t$ and $t+\delta t$.
If the fractional change in the population from one time to the next, $\delta
S/S(t)$, is a fixed value, the rate coefficient is independent of time and the
distance between $\ln S(t)$ and $\ln S(t+\delta t)$ is constant. We can
interpret the statistical distance as a criterion for the distinguishability of
$\ln S(t)$ at two times, since it is zero if the survival probability is time
independent, or as the square of the transition probability $k(t)\delta t$
during $\delta t$.

During an irreversible decay process from an initial time $t_i$ to a final time
$t_f$, and letting $\delta t\to 0$, we can integrate the arc length of the
logarithmic survival curve $\ln S(t)$
\begin{eqnarray}
  \mathcal{L}(\Delta t)
  = \int_{t_i}^{t_f}ds
  = \int_{t_i}^{t_f}\sqrt{\frac{I(t)}{S(t)}}\,dt
  = \int_{t_i}^{t_f}k(t)\,dt
\end{eqnarray}
to get the statistical length, $-\ln S(t)\big|_{S(t_i)}^{S(t_f)}$, measuring
the cumulative rate coefficient. The statistical length is infinite for an
infinite time interval $\Delta t=t_f-t_i$. Another useful quantity, related to
the statistical length, is the Fisher divergence, $\mathcal{J}(\Delta t)$. The
Fisher divergence we define as
\begin{eqnarray}
  \frac{\mathcal{J}(\Delta t)}{\Delta t}
  = \int_{t_i}^{t_f}ds^2
  = \int_{t_i}^{t_f}\frac{I(t)}{S(t)}\,dt
  = \int_{t_i}^{t_f}k(t)^2\,dt,
\end{eqnarray}
the time integrated square of the rate coefficient (times the magnitude of the
time interval). Both the statistical length and the Fisher divergence
are cumulative properties of the rate coefficient, identified through
Equation~\ref{eqn:fisherinrate} for irreversibly decaying populations. These
properties for the history of $k(t)$ satisfy an inequality
\begin{equation}
  \label{eqn:main}
  \mathcal{L}(\Delta t)^2 \leq \mathcal{J}(\Delta t),
\end{equation}
which is our main result. An analogous inequality in finite-time thermodynamics
is a bound on the dissipation in an irreversible
process~\cite{SalamonB83,SalamonNB84,FAQS85,Crooks07,FengC09,SivakC12}.

We will show the inequality above measures disorder in kinetic phenomena and is
an equality in the absence of both static or dynamic disorder. Some evidence
for this comes from the inequality $\mathcal{J}(\Delta t)-\mathcal{L}(\Delta
t)^2\geq 0$ in terms of the observed rate coefficient
\begin{eqnarray}
  \Delta t\int_{t_i}^{t_f}k(t)^2\,dt-\left[\int_{t_i}^{t_f}k(t)\,dt\right]^2
  &\geq&
  0.
\end{eqnarray}
For decay processes with a time independent rate coefficient, there is no
disorder, and the bound holds $\left(k\Delta t\right)^2 = k^2\Delta t^2$. While
this finding is suggestive, we can show more concretely that the inequality
measures the variation of the rate coefficient in irreversible decay kinetics.
We turn to this now and show the inequality measures the amount of disorder
during a given time interval. As a proof-of-principle we apply the theory to
widely used kinetic models for populations decaying non-exponentially. Note,
however, that the theory is model independent.

\section{Kinetic model with dynamic disorder}

\begin{figure}[t]
  \centering
  \includegraphics[width=0.98\columnwidth]{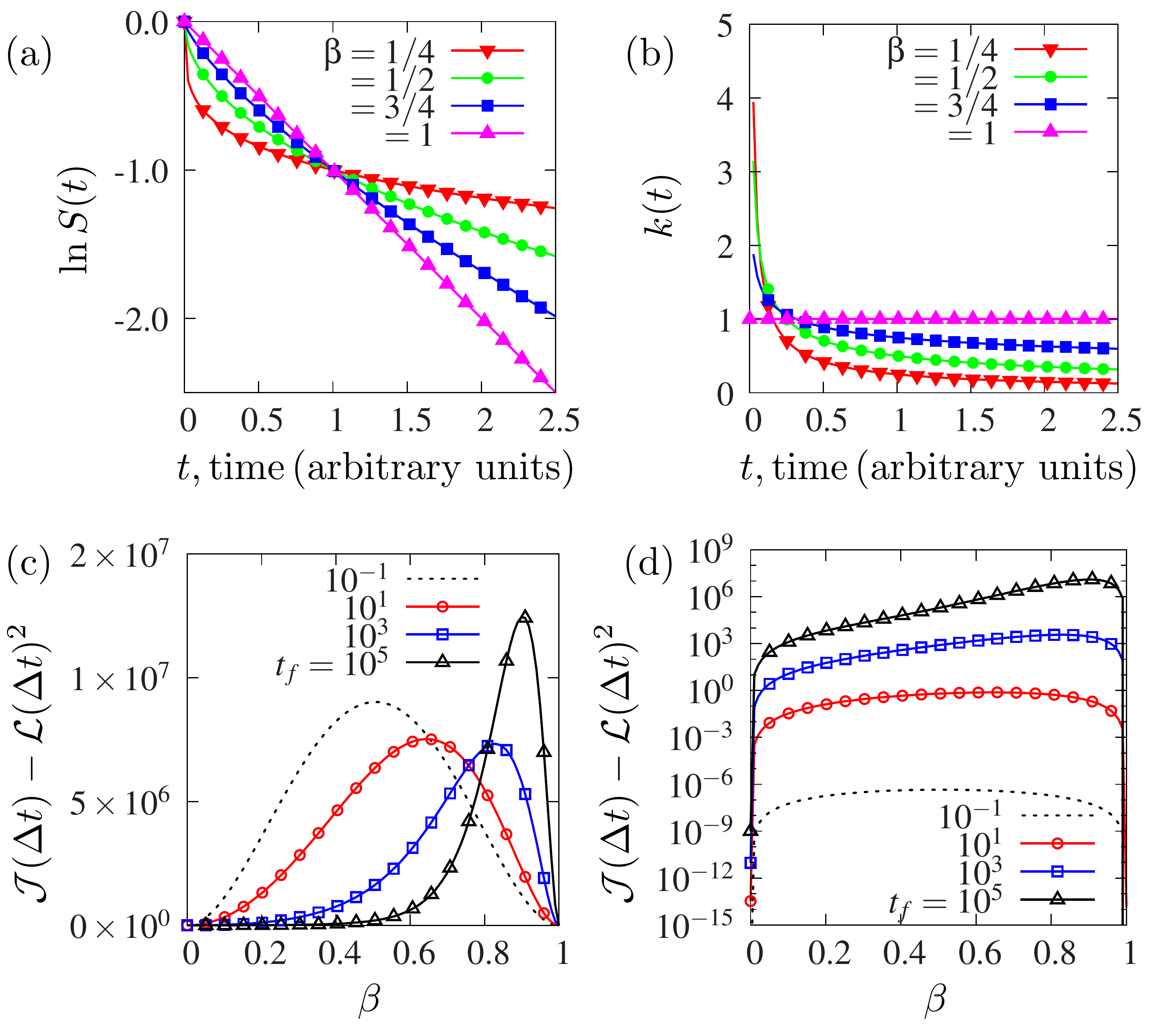}

  \caption{\label{fig:kww}(Color online)(a) Logarithmic
Kohlrausch-Williams-Watts survival curves $\ln S(t)$ for stretched exponential
decay with $\omega=1$ and $\beta=\tfrac{1}{4}$, $\tfrac{1}{2}$, $\tfrac{3}{4}$,
and exponential decay $\beta = 1$. (b) The rate of decay of the observed rate
coefficient $k(t)$ over time depends on $\beta$, $0<\beta\leq 1$. (c)
$\mathcal{J}(\Delta t)-\mathcal{L}(\Delta t)^2\geq 0$ over the range of $\beta$
values for the final times of $t_f = 10^{-1}$, $10^{1}$, $10^{3}$, $10^{5}$ and
an initial time of $t_i=1$, in arbitrary time units. The difference is scaled
by an arbitrary factor of $2\times 10^{13}$, $10^{7}$, $2\times 10^{3}$, and
$1$ for a direct comparison. (d) $\mathcal{J}(\Delta t)-\mathcal{L}(\Delta
t)^2$ versus the disorder parameter $\beta$ on a logarithmic scale.
$\mathcal{J}(\Delta t)=\mathcal{L}(\Delta t)^2$ for all $\beta$ and any time
interval when there is no dynamic disorder -- when $\beta = 1$ and $k=\omega$.}

\end{figure}

To demonstrate the inequality measures dynamic disorder, we consider the
Kohlrausch-Williams-Watts (KWW) stretched exponential survival function. This
empirical model is a widely used phenomenological description of relaxation in
complex media. It has been applied to the discharge of
capacitors~\cite{Kohlrausch54}, the dielectric spectra of
polymers~\cite{WilliamsW70}, and more recently the fluorescence of single
molecules~\cite{FlomenbomVLCEHRNVdK05}. The survival probability,
$\exp\left[-(\omega t)^\beta\right]$, has two adjustable parameters $\omega$ -
a characteristic, time-independent rate coefficient or inverse time scale - and
$0 < \beta \leq 1$ measuring the ``cooperativity'' of decay events - being
nearer or equal to $1$ for independent decay events and nearer $0$ for coupled
decay events (Figure~\ref{fig:kww}). Fitting data to this function when the
decay is non-exponential implicitly assumes the kinetics are dynamically
disordered.

Figure~\ref{fig:kww}(a) shows analytical results for how the natural logarithm
of the survival function versus time depends on the value of $\beta$. There is
a linear dependence on time only when $\beta=1$, corresponding to exponential
kinetics and a time independent rate coefficient $k(t)\to\omega$.  For all
other $\beta$ values the survival function decays faster than exponential
before, and slower than exponential after, $t=1/\omega=1$. These stretched
exponential decays have a time-dependent rate coefficient $k(t) =
\beta\left(\omega t\right)^\beta/t$ when $\beta\neq 1$ from
Figure~\ref{fig:kww}(b). The effective rate coefficient is the (negative of
the) slope of the graph of $\ln S(t)$ versus $t$. Through the rate law, the
decrease in $k(t)$ over time indicates a decrease in the rate of decay; this
also means the number of species decaying in a unit time interval is
decreasing. Furthermore, if we interpret the decay as an activated process over
an energy barrier~\cite{Zwanzig90}, a decrease in $k(t)$ implies an increase in
barrier height and slowing rate of escape. In the limit of long times, the rate
coefficient $k(t)$ is effectively constant, regardless of $\beta$. The value of
$\beta$ determines the rate at which $k(t)$ reaches this limit: smaller
$\beta$ values lead to a more rapid decline of $k(t)$.

From the effective rate coefficient, we find the statistical length
$\mathcal{L}_\textrm{KWW}(\Delta t)=\omega^{\beta} t^\beta\big|_{t_i}^{t_f}$
and, for $\beta\neq\tfrac{1}{2}$, the divergence
\begin{equation}
  \mathcal{J}_\textrm{KWW}(\Delta t) =
  \frac{\beta^2}{2\beta-1}\omega^{2\beta}\Delta t\,t^{2\beta-1}\big|_{t_i}^{t_f}.
\end{equation}
For $\beta=\tfrac{1}{2}$, the divergence is $\beta^2\omega^{2\beta}\Delta t\ln
t\big|_{t_i}^{t_f}$. From these results for stretched exponential kinetics of
the KWW type, we find $\mathcal{J}(\Delta t)\geq\mathcal{L}(\Delta t)^2$
(dropping the subscript KWW for clarity) and can show the inequality measures
the amount of dynamic disorder associated with the empirical parameter $\beta$.

The temporal variation of $k(t)$ on the observational time scale dictates the
magnitude of $\mathcal{J}(\Delta t)-\mathcal{L}(\Delta t)^2$. Since $\beta$
determines how strongly $k(t)$ varies in time, we show $\mathcal{J}(\Delta
t)-\mathcal{L}(\Delta t)^2$ over the range of $\beta$ for select final times in
Figure~\ref{fig:kww}(c) and (d). The plots are at four different final times
$t_f=10^{-1}$, $10$, $10^3$, $10^5$ and an initial time $t_i=1$ in arbitrary
units. Most striking in Figure~\ref{fig:kww}(c) and (d), is the maximum in
$\mathcal{J}(\Delta t)-\mathcal{L}(\Delta t)^2$. We see the maximum of the
inequality is at a $\beta$ greater than or equal to $0.5$ for all non-zero time
intervals and all $t_f$. At final times less than $1/\omega$, the difference is
symmetric with a maximum at $\beta=0.5$. With increasing $t_f$, the curve
becomes asymmetric and the maximum shifts to higher $\beta$ values.

In part, the maximum comes from $\mathcal{J}(\Delta t)-\mathcal{L}(\Delta t)^2$
being zero at the limiting values of $\beta$ where the effective rate
coefficient is time-independent. When $\beta=1$, the decay is exponential,
$k(t)=\omega$, and $\mathcal{J}(\Delta t)=\mathcal{L}(\Delta t)^2$ for all
final and initial times $t_f > t_i$. The effective rate coefficient is also
zero when $\beta=0$, though this limit is typically excluded from the KWW
model; however, when the rate coefficient is zero the equality again holds.
Since $\mathcal{J}(\Delta t)-\mathcal{L}(\Delta t)^2$ must be zero at the
limits of the $\beta$ range, and is non-zero between these limits, there must
be a maximum at intermediate $\beta$ values. In this case, and the model we
discuss next, there is a single maximum in $\mathcal{J}(\Delta
t)-\mathcal{L}(\Delta t)^2$ versus $\beta$. A single maximum results when the
decay is monotonic, and non-exponential, with a time-varying rate coefficient
between two limits where the rate coefficient is constant. However, if the
decay is more complex, or if the kinetics are driven, more maxima are possible.

The maximum in $\mathcal{J}(\Delta t)-\mathcal{L}(\Delta t)^2$ shown in
Figure~\ref{fig:kww}(c) and (d) is also the result of how the history of $k(t)$
depends on $\beta$ in this model. Panel (b) shows that decreasing $\beta$ also
increases the difference between $k(t_f)$ and $k(t_i)$, which would imply an
increase in the difference between the barrier heights at $t_f$ and $t_i$ for an
activated process. As $\beta$ decreases from one to zero, this increase in the
variation of the rate coefficient initially leads to a greater inequality
between $\mathcal{J}(\Delta t)$ and $\mathcal{L}(\Delta t)^2$.  However, as
$\beta$ continues to decrease, the effective rate coefficient also becomes a
more rapidly decreasing function of time (Figure~\ref{fig:kww}(b)).
Consequently, significant changes in the magnitude of $k(t)$ can become
relatively small portions of the observational time window. For example, at
$t_f=10^5$ the effective rate coefficient varies little over the majority of
the observation time for $\beta<0.5$, and $\mathcal{J}(\Delta
t)\approx\mathcal{L}(\Delta t)^2$.  But, for $\beta>0.5$, $k(t)$ is more
strongly dependent on time over the interval $t_f-t_i$ and $\mathcal{J}(\Delta
t)>\mathcal{L}(\Delta t)^2$. Overall, the magnitude of the $\mathcal{J}(\Delta
t)-\mathcal{L}(\Delta t)^2$ reflects the both the variation of $k(t)$ in time
and the fraction of the observation time where those changes occur. The maximum
in the inequality versus $\beta$ reflects the competition of these factors.


\section{Kinetic model with static disorder}

\begin{figure}[t]
  \centering
  \includegraphics[width=0.98\columnwidth]{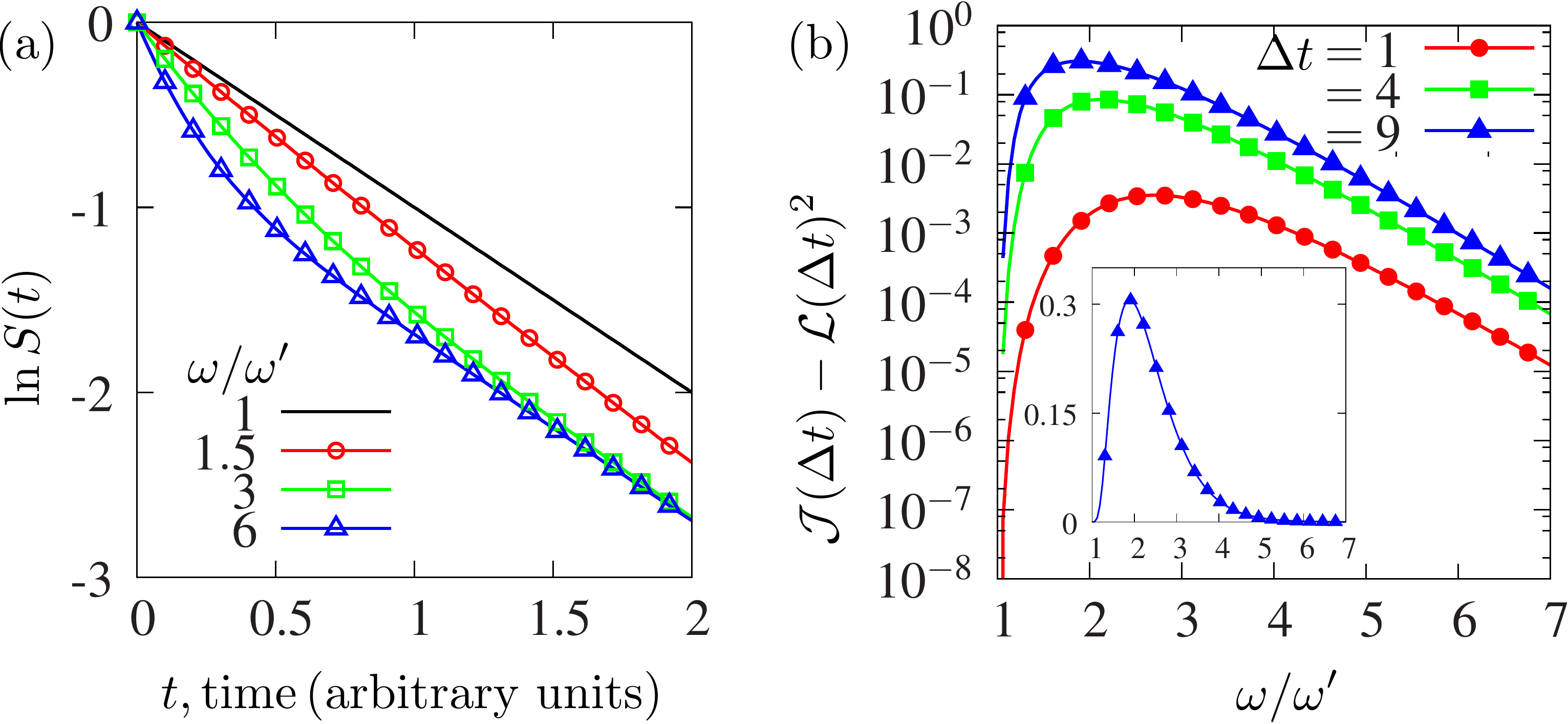}

  \caption{\label{fig:plonka}(Color online)(a) Logarithmic survival plot for
the kinetic scheme in the text for $a=1$ and $\omega'=1$. The decay kinetics
are statically disordered when $\omega\neq\omega'$, leading to the
non-exponential relaxation of $S(t)$ and the non-linearity of $\ln S(t)$ in
time. The kinetics are exponential when $\omega=\omega'$ and $\ln S(t)$ versus
$t$ is linear (black line). There is no disorder when $\omega\neq\omega'$ at
long times when the slower channel dominates decay. (b) The graph of
$\mathcal{J}(\Delta t)-\mathcal{L}(\Delta t)^2$ reflects these facts, showing
this difference measures the amount of static disorder over a range of relative
decay rates $\omega/\omega'$ ($\omega'=1$) and time intervals $\Delta t=1$,
$4$, and $9$ on a logarithmic scale. The inset is on a linear scale for $\Delta
t=9$. $\mathcal{J}(\Delta t)-\mathcal{L}(\Delta t)^2$ is zero if decay is
exponential (when $\omega=\omega'$ or when $\omega\gg\omega'$), but nonzero
otherwise.}

\end{figure}

The difference between $\mathcal{J}(\Delta t)$ and $\mathcal{L}^2(\Delta t)$
also measures the amount of static disorder in irreversible decay kinetics over
a time interval. As a tractable example, consider the model with two
experimentally indistinguishable states, $A$ and $A'$~\cite{GoldanskiiKT97}.
\begin{figure}[!ht]
  \centering
  \includegraphics[width=0.90in]{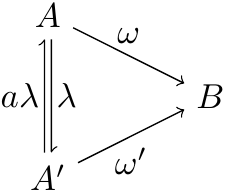}
\end{figure}
These states might be the internal configurations of a reactive molecule that
can interconvert through isomerization or two energy levels leading to
unimolecular dissociation in the gas phase. The states are in equilibrium and
so a population occupying these states will undergo transitions between them. A
population occupying the $A$ and $A'$ states can also decay irreversibly to $B$
with rates characterized by two time-independent rate coefficients, $\omega$
and $\omega'$. If the individual decay out of $A$ and $A'$ have different
characteristic rates, it may be necessary to characterize the collective decay
from the $A$ states with both rate coefficients. Since only two rate
coefficients $\omega$ and $\omega'$ represent the ``fluctuations'' in the rate
coefficient, this is the simplest case of potentially statically disordered
kinetics (Figure~\ref{fig:plonka}), and is readily generalizable to a continuum
of $A$ states.

The phenomenological rate equations are
\begin{eqnarray}\nonumber
  \der{N_A}{t} &=& -(\omega + \lambda)N_A + a\lambda N_{A'}\\[7pt]\nonumber
  \der{N_{A'}}{t} &=& -(\omega' + a\lambda)N_{A'} + \lambda N_A
\end{eqnarray}
for the number of species in state $A$, $N_A$, and $A'$, $N_{A'}$.  These
numbers are exact and not fluctuating in time. Whether the decay from $A$ and
$A'$ is exponential or non-exponential depends on the relative rates of
interconversion and reaction. Two limiting solutions to these rate equations
are of particular interest~\cite{Plonka01}. Each limit affects the analytical
form of the survival probability for the population of the $A$ and $A'$ states
\begin{equation}
  S(t) = \frac{N_A(t) + N_{A'}(t)}{N_A(0) + N_{A'}(0)}.
\end{equation}

First, in the limit where interconversion between $A$ and $A'$ is
fast compared to the reaction, $\lambda\to\infty\gg \omega,\omega'$, the
survival probability is
\begin{equation}
  S(t) \sim
  \exp{\left[-\left(\frac{a}{1 + a}\omega + \frac{1}{1 + a}\omega'\right)t\right]}.
\end{equation}
and the population decays exponentially. From Equation~\ref{eqn:key},
the observed rate coefficient $k(t)$ is time independent and a weighted-average
of $\omega$ and $\omega'$, $\tfrac{a}{1+a}\omega+\tfrac{1}{1+a}\omega'$. From
this limiting survival probability it follows that the divergence and length
squared are equal, $\mathcal{L}^2(\Delta t) = \mathcal{J}(\Delta t) =
k(t)^2\Delta t^2$. This result illustrates that to satisfy the bound, the $A$
states must be kinetically indistinguishable, which happens here because the
population collectively decays to products from one effective state.

Second, in the limit $\lambda\to 0$, the transitions between the $A$
states are slow relative to those of reaction. When the decay from $A$ and
$A'$ is independent, the net decay is bi-exponential
\begin{equation}
  \label{eqn:s}
  S(t) \sim \frac{a}{1 + a}e^{-\omega t} + \frac{1}{1 + a}e^{-\omega't}.
\end{equation}
The logarithmic dependence of this survival function versus time in
Figure~\ref{fig:plonka}(a) is linear, and the decay of $S(t)$ is effectively
exponential, if the decay from the $A$ and $A'$ states have the same rate
coefficients $\omega=\omega'$. In this case, we again find
$\mathcal{L}^2(\Delta t) = \mathcal{J}(\Delta t)$. However, this condition
breaks when the kinetics are statically disordered $\omega\neq\omega'$ and
decay from $A$ and $A'$ competitively depletes the population, leading to the
non-exponential decay in Figure~\ref{fig:plonka}(a). The $S(t)$ in this regime
causes the observed rate coefficient $k(t)$ to vary in time. Only when there is
static disorder in this model does $\mathcal{L}(\Delta t)^2$ not equal
$\mathcal{J}(\Delta t)$.

The relative magnitude of $\omega$ and $\omega'$, which we measure with the
ratio $\omega/\omega'$, and the time interval, $\Delta t$, control the degree
of static disorder. The difference $\mathcal{J}(\Delta t)-\mathcal{L}(\Delta
t)^2$ is sensitive to both of these factors. Figure~\ref{fig:plonka}(b)
illustrates the dependence of the inequality on $\omega/\omega'$ and assumes
the survival function is valid, $\omega,\omega'\ll \lambda$, over the range of
interest. Varying $\omega/\omega'$ shows the difference between the divergence
and length squared is positive, has a maximum that depends on the time
interval, and tends to zero with increasing $\omega/\omega'$. In the figure,
the difference is shown for three final times $t_f=2$, $5$, and $10$ and
$t_i=1$, in arbitrary units. A logarithmic scale allows direct comparison of
these curves (the vertical axis in the inset has a linear scale).

An inequality between $\mathcal{L}(\Delta t)^2$ and $\mathcal{J}(\Delta t)$
gives quantitative insight into the mechanism of kinetic processes, when only
the survival of the entire population is available. From
Figure~\ref{fig:plonka}(b) the inequality grows but as the disparity in the
rate coefficients of each decay path increases further, there becomes
a separation of timescales, one path decays on times that are increasingly
short in comparison to the time interval of interest. Just as for the KWW
model, there is a maximum in the inequality between the divergence and length
squared. In the limit $\omega\gg\omega'$, $\mathcal{J}(\Delta
t)=\mathcal{L}(\Delta t)^2$, indicating exponential decay and a single
characteristic rate coefficient. The maximum is understandable from the
survival function: with increasing $\omega$ or $t$, the term $e^{-\omega t}$
decreases until, ultimately, the decay channel contributes negligibly to the
overall decay of the population -- the kinetics are exponential with the
characteristic rate of the slower decay channel, $k(t)\to\omega'=1$. The
survival function in Equation~\ref{eqn:s} becomes $S(t)\sim
(1+a)^{-1}\exp{(-\omega't)}$.

Together these findings demonstrate the inequality is a statement about whether
a single rate coefficient is sufficient for irreversible rate processes. The
rate coefficient is a unique constant only when $\mathcal{J}(\Delta
t)=\mathcal{L}(\Delta t)^2$, and the decaying states are kinetically
indistinguishable. Otherwise, the inequality between $\mathcal{J}(\Delta t)$
and $\mathcal{L}(\Delta t)^2$ measures the dispersion in the rate
coefficient--though it does not distinguish between statically and dynamically
disordered kinetics. The inequality both results from, and measures, the
heterogeneity of microscopic environments that ultimately produce disordered
kinetics. For activated processes over an energy barrier, heterogeneous
environments can cause either fluctuating barrier heights or a distribution of
barriers through steric/energetic effects. Consequently, the inequality could
be used to investigate the local structure around reaction sites in disordered
media and the dependence of rate processes on energetic or structural effects
of the environment~\cite{SiebrandW96}. The equality signifies when the
interconversion between decaying states is faster than decay, when there is no
microscopic heterogeneity to cause a distribution of rate coefficients, and when
observations of the process are on a time scale where fast decay processes are
insignificant to the longer term survival of the population. Under these
circumstances traditional kinetics holds.

\section{Conclusions}

In summary, the inequality between $\mathcal{L}^2(\Delta t)$ and
$\mathcal{J}(\Delta t)$ measures the amount of static and dynamic disorder in
irreversible decay kinetics. The inequality follows from the relation between
the rate coefficient for decay and the Fisher information. It relies on two
functions of the Fisher information - a statistical length and the Fisher
divergence for the decaying population - which are properties of the history of
fluctuations in the rate coefficient. This relationship is a quantitative
signature of when decay kinetics are accurately described by a fluctuating rate
coefficient (measured by the inequality) and when a single, unique rate
constant is sufficient. Traditional kinetics corresponds to the condition
$\mathcal{L}^2(\Delta t)=\mathcal{J}(\Delta t)$. Our results suggest minimizing
the difference between $\mathcal{L}^2(\Delta t)$ and $\mathcal{J}(\Delta t)$,
when it is desirable to minimize the statistical description of kinetic
phenomena with rate coefficients or to maximize the predictive fidelity of rate
coefficients extracted from experimental or simulation data. In the future,
this framework may also be useful in the analysis of, potentially complex,
chemical reactions with fluctuating rates.

\section{Acknowledgements}

This work was supported as part of the Non-Equilibrium Energy Research Center
(NERC), an Energy Frontier Research Center funded by the U.S. Department of
Energy, Office of Science, Basic Energy Sciences under Award \#DE-SC0000989.
This material is based upon work supported in part by the U.S. Army Research
Laboratory and the U.S. Army Research Office under grant number
W911NF-14-1-0359.


%

\end{document}